\documentclass[aps,prl,twocolumn,amsmath,amssymb,superscriptaddress,floatfix,showpacs]{revtex4}

\usepackage{graphicx}
\usepackage{bm}
\input epsf.sty

\begin{document}


\title{Fluctuation Ratios in the Absence of Microscopic Time Reversibility}

\author{Sven Dorosz and Michel Pleimling}
 \affiliation{Department of Physics, Virginia Polytechnic Institute and State University, Blacksburg, Viriginia 24061-0435, USA}
 
\date{\today}

\begin{abstract}
We study fluctuations in diffusion-limited reaction systems driven out of their stationary state.
Using a numerically exact method, we investigate fluctuation ratios in various systems which differ by
their level of violation of microscopic time reversibility. 
Studying a quantity that for an equilibrium system is related to the work done to the system,
we observe that under certain conditions
oscillations appear on top of an exponential behavior of transient fluctuation
ratios. We argue that these oscillations  
encode properties of the probability currents in state space.

\end{abstract}


\maketitle

In recent years the study of fluctuations in nonequilibrium small systems has evolved into a
very active field of research,
see, e.g., \cite{Eva93,Eva94,Gal95,Jar97,Kur98,Leb99,Cro99,Cro00,Hat01,
Muk03,Sei05,Lip02,Col05,Dou05a,Wan02,Car04,Tie06,Ber08}. Various fluctuation and work theorems
have been formulated and their applicability has been verified in recent experiments
\cite{Lip02,Col05,Dou05a,Wan02,Car04,Tie06,Ber08}, demonstrating their usefulness
for characterizing out-of-equilibrium systems.
It is remarkable that these fluctuation relations
yield very generic statements valid for large classes of nonequilibrium systems.

Diffusion-limited systems with irreversible reactions form an important
class of systems that have not been studied thoroughly
in the context of fluctuation relations. 
In the past all discussed extensions of fluctuation theorems to nonequilibrium systems with chemical reactions
\cite{Gas04,Sei04,And04,Sei05a,And06,Sch07,And08} focused on reversible reactions and
reaction networks. Effectively, however, irreversible reactions can be encountered
if the products of the reactions are evacuated rapidly enough. What makes irreversible reactions so
interesting is that there is a major qualitative difference 
with reversible reactions: whereas in
the latter case microscopic time reversibility holds, in irreversible reaction-diffusion
systems microscopic time reversibility is usually broken. As we show in this Letter the
absence of microscopic reversibility leads to unexpected and nontrivial modifications
of the properties of transient fluctuations.

Systems with broken microscopic time reversibility are readily
found in granular matter. It has been claimed \cite{Fei04} that fluctuations in fluidized
granular medium are in accord with the Gallavotti and Cohen fluctuation theorem. As the
fluctuation theorem requires microscopic reversibility, this interpretation of the experimental
data is problematic and has been criticized \cite{Pug05}. However, in \cite{Bon06} it has been proposed
that under certain assumptions and for a specific timescale a fluctuation relation should be 
recovered in granular materials. In our study we will not be able to contribute directly to
this controversy, but the results presented in this Letter clearly show the interesting and nontrivial character
of fluctuations in systems in which microscopic reversibility is absent.

In diffusion-limited reaction systems the stationary states
can be true nonequilibrium states. Due to their relative simplicity, in conjunction with
a highly non-trivial physical behavior, reaction-diffusion systems are
considered to be paradigmatic examples of nonequilibrium many-body systems. Thus our current
understanding of nonequilibrium phase transitions \cite{Hin00} and of aging phenomena in absence of detailed
balance \cite{Hen07} has mainly emerged through numerous studies of the out-of-equilibrium behavior of these systems.

We consider here one-dimensional lattices of $N$ sites with periodic boundary conditions.
Forbidding multiple occupancy of a given lattice site, particles $A$ jump to unoccupied nearest neighbor
sites with a diffusion rate $D$ and undergo various reactions. We discuss in the following three different reaction
schemes, see Table \ref{table1}, and we denote with model 1, 2, and 3 the three models that result from these reaction
schemes. Obviously, the reactions change the number of particles in the system, whereas the diffusion keeps the
particle number constant.

\begin{table}[thb]
\begin{tabular}{|c|c|c|}
\hline
 model 1 & model 2 & model 3\\
\hline
$A+A\stackrel{\lambda}{\to} 0+A$ & $A+A\stackrel{\lambda}{\to} 0+A$ & $A+A\stackrel{\lambda}{\to} 0+0$\\
$A+0 \stackrel{h}{\to} A+A $& $0\stackrel{h}{\to} A $ & $0\stackrel{h}{\to} A $\\
\hline
\end{tabular}
\caption{The three reaction schemes discussed in this Letter. A new particle can only be created at an empty 
lattice site.}\label{table1}
\end{table}

For fixed values of the reaction and diffusion rates, model 1 is in (chemical) equilibrium. This is different
for the other two models where microscopic reversibility is partly or fully broken. By breaking microscopic
reversibility we mean that
if $\omega(C \longrightarrow C')$ is the transition probability from configuration $C$ to configuration $C'$,
we can have the situation that $\omega(C \longrightarrow C') = 0$ even though $\omega(C' \longrightarrow C) > 0$. 
For model 2 we observe that some reactions are reversible whereas others are not. For example,
whereas we can create a new particle in the middle of two empty sites, $000 \longrightarrow 0A0$ with rate $h$,
it is not possible to directly go back to the configuration with three empty sites by 
destroying this isolated $A$ particle,
as we need to have two neighboring $A$ particles for particle annihilation. This is different for
$00A \longrightarrow 0AA$, as here a direct path back to the initial configuration exists. Finally, in model 3
microscopic reversibility is broken for all reactions.

\begin{figure}[h]
\centerline{\epsfxsize=3.30in\ \epsfbox{
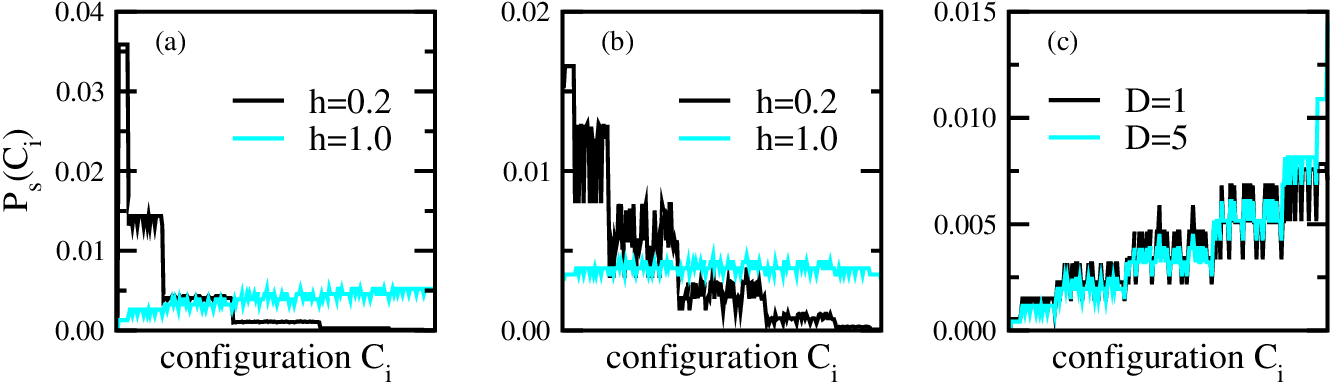}}
\caption{(Color online) Stationary probability distributions for (a) model 1 with $\lambda=1$ and $D=1$, (b) model 3
with $\lambda=1$ and $D=1$, and (c) model 2 with $\lambda=1$ and $h=1.4$. 
Shown are the distributions for two values of the creation rate $h$ respectively of the diffusion constant $D$
for systems with $N = 8$ lattice 
sites. The configurations are grouped by number of particles.
}
\label{fig1}
\end{figure}

We can readily access the stationary probability distributions,
i.e. the probabilities $P_s(C_i)$ to encounter the microscopic configuration 
$C_i$ in a given stationary state.
This is done in the usual way by rewriting the master equations in matrix form involving the Liouvillian and
by noticing that the stationary probabilities form the unique eigenvector of this operator to the eigenvalue 0.
With $N$ sites we have $2^N$ configurations as there is at most one particle on each lattice site. 
The null eigenvector of the resulting $2^N \times 2^N$ matrix is obtained using standard 
algorithms. Figure \ref{fig1} shows the stationary probability distributions for three different cases.
Configurations with the same number of particles are grouped, with the empty
configuration to the left and the fully occupied lattice to the right. When the creation of new particles takes place
with a small rate, see Fig. \ref{fig1}a and \ref{fig1}b, 
the most probable configurations are those with few occupied sites, whereas the configurations
with more occupied sites have an increasing weight for increasing creation rates. Similar changes are observed
when changing the rate $\lambda$. Obviously, a change of reaction rates has a large impact on the 
stationary probability distributions. This is different for the diffusion constant $D$ which only changes
the distributions quantitatively and not qualitatively, as shown in Fig. \ref{fig1}c.
We remark that even though there are 
visible differences in the stationary probability distributions, 
it is far from obvious how one should infer from these distributions
the equilibrium (model 1) or strongly nonequilibrium (model 3) nature of the system. 

In order to get a better understanding of our systems we look at the transient behavior when we
drive the system from one stationary state to another by changing a
reaction rate. Experimentally, a change of rates in chemical reactions 
can be achieved by changing the temperature.
In our protocol we change one of the rates $r$ from an initial value $r_0$ to a final value $r_M$ in $M$ 
equidistant steps of length $\Delta r$, yielding the values
$r_i = r_0 + i \Delta r$ with $i=0,\cdots , M$.
We compute the observable \cite{Hat01}
\begin{equation} \label{dp}
\delta \phi = \sum\limits_{i=0}^{M-1} \left( \ln P_s(C_i,r_{i+1}) - \ln P_s(C_i,r_i) \right)
\end{equation}
where $P_s(C_i,h_i)$ is the probability to find the configuration $C_i$ in the stationary state
corresponding to the value $r_i$ of the reaction rate $r$. For a system in thermal equilibrium
the quantity $\delta \phi$ is given by $\delta \phi = \beta (W - \Delta F)$, where $\beta$ is the
inverse temperature, $W$ is the work done to the system, and $\Delta F$ is the difference between
the free energies of the initial and final states. It is important to note that the quantity
(\ref{dp}) is still well defined in absence of microscopic reversibility. This is not the case for
many of the quantities that have been studied recently in the context of fluctuation relations.

Hatano and Sasa \cite{Hat01} proved for Langevin systems with continuous dynamics that the quantity (\ref{dp})
fulfills in the limit $M \longrightarrow \infty$ the following simple fluctuation relation:
\begin{equation} \label{hs}
\langle e^{-\delta \phi} \rangle =1
\end{equation}
where the average is the average over all possible histories relating the initial and final steady states.
For an equilibrium system the relation (\ref{hs}) reduces to the well-known Jarzynski relation  \cite{Jar97}.
Even though not explicitly stated in \cite{Hat01}, the property (\ref{hs}) 
of $\delta \phi$ can be shown in a straightforward
way to also hold in systems with discrete dynamics, and this independently on whether
microscopic reversibility prevails or not. The verification of the relation (\ref{hs}) is therefore a very good
benchmark in order to validate our numerical approach.

Changing the rate $r$ from the initial value $r_0$ to 
the final value $r_M$ in $M$ steps, we can easily compute the exact 
stationary probability distributions for any value $r_i$ with
$i=0,\cdots , M$. In order to verify Eq. (\ref{hs}) we need to generate all possible sequences
of configurations ({\it paths} in configuration space) $C_0 \longrightarrow C_1 \longrightarrow \cdots 
\longrightarrow C_{M-1} \longrightarrow C_M$, determine the weights $\prod\limits_{i=0}^M P_S(C_i,r_i)
\omega(C_i \longrightarrow C_{i+1}, r_{i+1})$
and the values of $\delta \phi$ along the different paths, and average over all these possibilities. 
Here $\omega(C_i \longrightarrow C_{i+1}, r_{i+1}$ is
the transition probability from configuration $C_i$ to configuration $C_{i+1}$ at the value $r_{i+1}$
of our reaction rate.
With this numerical exact calculation we verify for all studied cases 
the validity of the integral fluctuation relation (\ref{hs})
with deviations less than $10^{-7}$.

As in our numerically exact approach we generate all paths recursively,
the CPU time needed for the generation of all trajectories 
grows exponentially with the lattice size $N$ and the number of steps $M$, and only rather
small system sizes (with $N < 10$) can be accessed in this way. 
For example, for $N =8$  we generate $2.6~10^8$ different trajectories for $M=6$, whereas
$2.7~10^{10}$ trajectories are generated for $M=8$.
We also studied larger systems through Monte Carlo simulations and checked
that these results are consistent with the numerical exact results obtained for the small systems.
For this reason we focus in the following on the numerically exact results and defer a discussion
of the Monte Carlo simulations to later \cite{Dor09}.

\begin{figure}[h]
\centerline{\epsfxsize=3.30in\ \epsfbox{
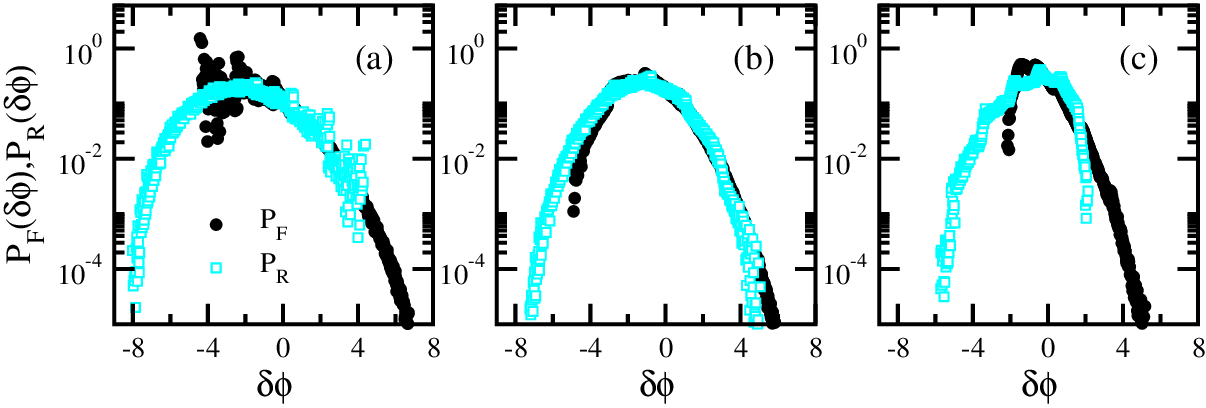}}
\caption{(Color online) Probability distributions $P_F$ and $P_R$ as a function of $\delta \phi$ for the forward and reverse processes:
(a) model 1, (b) model 2, (c) model 3, with $\lambda =1$, $D=1$,  and $N=8$.
In all cases the rate of particle creation was changed from $h_0=0.2$ to $h_M = 1.4$ in $M=8$ steps.
The scattering in the data is not due to poor statistics, as we are using a numerical exact method
for the computation of the probability distributions.
}
\label{fig2}
\end{figure}

\begin{figure}[h]
\centerline{\epsfxsize=3.30in\ \epsfbox{
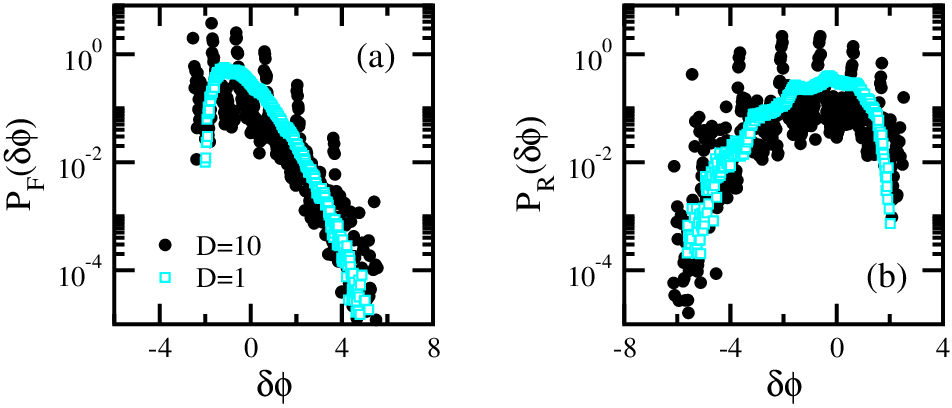}}
\caption{(Color online) Probability distributions $P_F$ (a) and $P_R$ (b) as a function of $\delta \phi$ for model 3
with $\lambda =1$, $N=8$, and $h$ changing from $h_0=0.2$ to $h_M = 1.4$ in $M=6$ steps.
The peaks are more pronounced for larger values of $D$.
}
\label{fig3}
\end{figure}

%
Before discussing the detailed fluctuation relation, let us first 
look at the probability distribution $P_F(\delta \phi)$ of the
quantity $\delta \phi$ for the forward process where the rate $r$ is changed from $r_0$ to $r_M$ 
as well as at the probability distribution $P_R(\delta \phi)$ for the reversed process where $r$ is
changed from $r_M$ to $r_0$. In the reversed process the rate $r$ takes on 
the same values as in the 
forward process but in the reversed order. We show in Figure \ref{fig2} the resulting probability
distributions for the three models with $N=8$ sites where we change the creation rates from $h_0=0.2$
to $h_M = 1.4$ in $M = 8$ equidistant steps. Interestingly, the probability distributions 
are skewed distributions that exhibit additional intriguing peaks.
Increasing the diffusion rate $D$ leads to a sharpening of these peaks, as is shown in 
Figure \ref{fig3} for model 3 with $M=6$ and different values of $D$.
We checked that the main contributions to these peaks comes from those trajectories in configuration space
where diffusion steps abound, whereas reactions,
which change the number of particles in the system, only take place rarely.
It should be noted that for very large values of $D$ the peaks also appear for the equilibrium model 1 and are
therefore not characteristic of broken microscopic time reversibility.

\begin{figure}[h]
\centerline{\epsfxsize=3.30in\ \epsfbox{
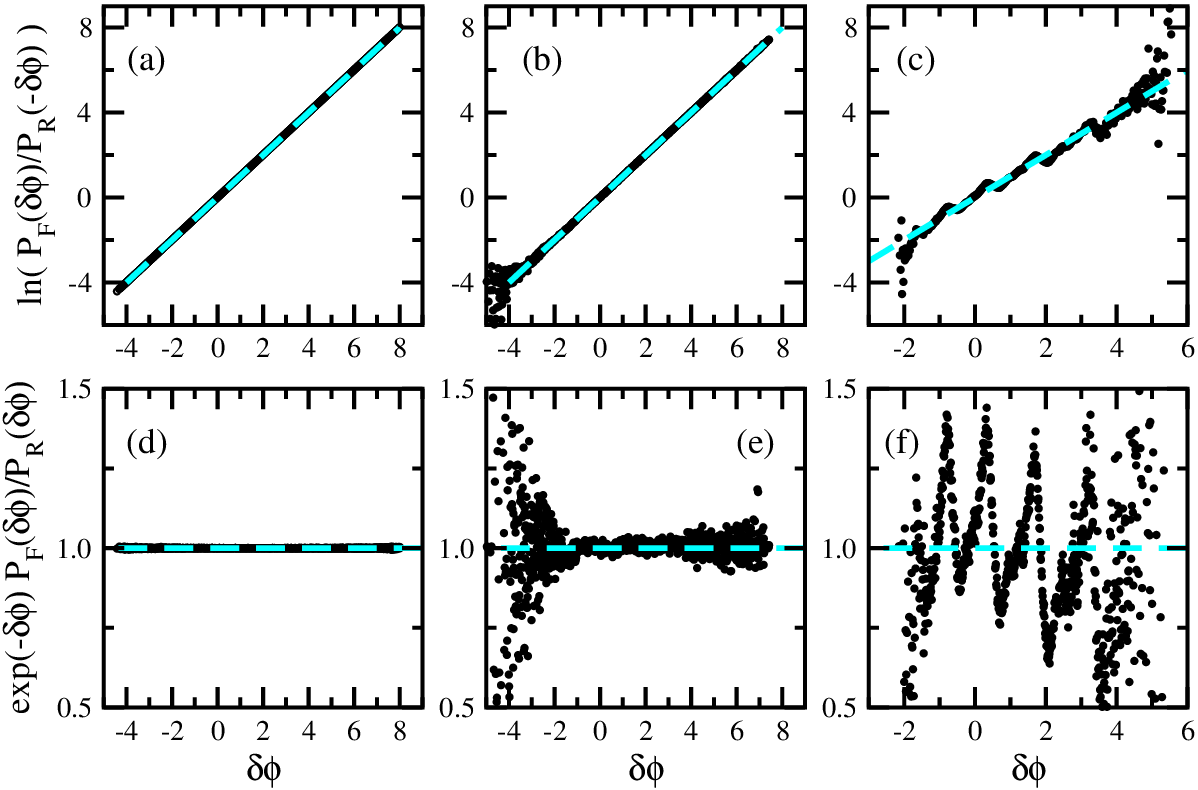}}
\caption{(Color online) Ratio $P_F(\delta \phi)/P_R(-\delta \phi)$ as a function of $\delta \phi$ for (a) and (d) model 1,
(b) and (e) model 2, and (c) and (f) model 3, with $\lambda =1$, $D=1$, $N=8$, and $M =8$, whereas 
$h$ is changed from $h_0=0.2$ to $h_M = 1.4$. In the lower panels we plot $e^{-\delta \phi} 
P_F(\delta \phi)/P_R(-\delta \phi)$ in order to highlight deviations from the equilibrium behavior 
(\ref{pfdpr}). The dashed lines indicate the expected behavior (\ref{pfdpr}) for a system in equilibrium
before and after the change of parameter.
}
\label{fig4}
\end{figure}

In Figure \ref{fig4} we discuss the fluctuation ratio $P_F(\delta \phi)/P_R(-\delta \phi)$ for the
observable (\ref{dp}). For a system that fulfills detailed balance for all values of 
the rate $r$ we expect that
\begin{equation} \label{pfdpr}
P_F(\delta \phi)/P_R(-\delta \phi) = \exp( \delta \phi )~.
\end{equation}
Indeed, it is straightforward to show that for a system initially in thermal equilibrium relation (\ref{pfdpr})
together with the definition (\ref{dp}) of the quantity $\delta \phi$ yields the Crooks relation
\begin{equation} \label{crooks}
P_F(W)/P_R(-W) = \exp\left[\beta(W - \Delta F ) \right]~.
\end{equation}
In order to highlight any deviations from the exponential behavior, we plot in the lower panels of Fig. \ref{fig4}
the quantity
$e^{-\delta \phi} 
P_F(\delta \phi)/P_R(-\delta \phi)$.
Looking at Fig. \ref{fig4}a and \ref{fig4}d, we observe that for the equilibrium model 1
the ratio of the two probability distributions indeed displays a perfect exponential behavior. As the
probability distributions themselves are skewed distributions, see Fig. \ref{fig2}, this is a 
nontrivial result that nicely demonstrates the importance of the Crooks relation.

As already discussed, microscopic reversibility is partly broken for model 2: whereas many trajectories
in configuration space are fully reversible, this does not hold true for all of them.
We observe that the probability distribution ratio still displays an exponential behavior on average, see Fig. \ref{fig4}b,
but the data do not fall any more exactly on the exponential curve but instead are scattered around that curve 
(Fig. \ref{fig4}e). 

For model 3, where microscopic reversibility is absent, 
a remarkable change takes place and systematic deviations from the exponential behavior are observed, see
Fig. \ref{fig4}c and \ref{fig4}f. These deviations take the form of oscillations. As we argue in the
following, these deviations reveal properties
of the probability currents in state space.

In order to develop a better understanding for the origin of these oscillations, we 
studied systematically the dependence of this feature on the reaction and diffusion rates as well as on the
system size and the number $M$ of elementary steps \cite{Dor09}. In fact, the oscillations are very robust and
are encountered for all studied values of the system parameters. We also observe that a
change of the positions of the peaks is 
directly related to a qualitative change of the stationary probability distributions, as the position of the
peaks strongly shift when the reaction rates are changed, but do only change slightly when the diffusion constant
is modified. Changing the diffusion constant, however, greatly enhances the peak height.

At this stage one might think that the peaks observed in the probabilities $P_F$ and $P_R$, see Fig.
\ref{fig2} and \ref{fig3}, are the origin of the peaks in the probability ratio. On the one hand,
there is of course an intimate relation between the peaks in $P_F$ and $P_R$ and those encountered when taking
the ratio of these two probabilities. On the other hand, however, peaks also appear in $P_F$ and $P_R$ for models
2 and 1, even though no peaks are observed for the corresponding ratio. 
The appearance of peaks in the probabilities $P_F$ and $P_R$
is therefore a necessary condition, but it alone can not explain our observations.

It is important to note that model 3 differs qualitatively from models 1 and 2.
For all the models the configuration
space is divided into different subspaces, characterized by a constant number of particles in the system,
which are invariant under the action of diffusion. A passage from
one subspace to the other only takes place when the number of particles is changed by a reaction.
In models 1 and 2 every reaction changes the number of particles by one, thus connecting
different subspaces pairwise. One of the consequences of this is
that the peaks in the distributions $P_F$ and $P_R$ compensate each other when computing the ratio
$P_F(\delta \phi)/P_R(-\delta \phi)$. This compensation is only approximate for model 2 due to the fact that
some trajectories can not be travelled in the reversed direction when reversing the protocol. The situation
is different for model 3 as here we have an asymmetry in the change of particle numbers:
whereas the number of particles is enhanced by one in the creation process,
two particles are always destroyed in the annihilation process.
Consequently, the trajectories in configuration space for the forward and backward processes are completely different,
as they connect the different subspaces with fixed number of particles in a different way. 
It follows that probability currents for the
forward and backward process are also very different. This
yields probability distributions $P_F$ and $P_R$ whose peaks do not compensate each other when the ratio
is formed, thus giving place to the observed systematic deviations. It is clear from this discussion that we
expect this mechanism, and therefore the observed systematic deviations,
to be common in systems where the absence of microscopic reversibility is accompanied by an asymmetry in the probability
currents in configuration space.

In summary, we have studied reaction-diffusion systems and showed that the absence of microscopic reversibility
can lead for transient fluctuation ratios for the observable $\delta \phi$
to systematic deviations from the exponential behavior 
encountered in systems with equilibrium steady states. These deviations take the form of oscillations,
and we argue that this intriguing feature reveals properties of the probability
currents in state space.

It is a pleasure to thank Chris Jarzynski, Uwe T\"{a}uber, Fr\'{e}d\'{e}ric van Wijland and Royce Zia
for interesting and helpful discussions.

\end{document}